\documentstyle[aps]{revtex} 
\input{psfig.sty}
\title{Baryons in O(4) and Vibron Model}
\author{ M.\ Kirchbach$^1$,  M.\ Moshinsky$^2$, Yu.\ F.\ Smirnov$^3$ }
\address{
$^1$Facultad  de F\'isica, Universidad Aut\'onoma de Zacatecas, \\ 
        A.\ P.\ 600, ZAC-98062, M\'exico\\
$^2$ Instituto de F\'isica, 
Universidad Nacional Aut\'onoma de M\'exico\\
A.\ P.\ 20-364, 01000 M\'exico, D.\ F.\ , M\'exico\\
$^3$ Instituto de Ciencias Nucleares, Universidad Nacional Aut\'onoma de
M\'exico,\\
A.\ P.\ 70-543, 04510 M\'exico, D.\ F.\ , M\'exico}
\begin{document}

\maketitle

\begin{abstract}
The structure of the reported excitation spectra of the light 
unflavored baryons is described in terms of multi-spin valued 
Lorentz group representations of the so called Rarita-Schwinger (RS) type
$\left({K\over 2}, {K\over 2}\right)\otimes \left[
\left({1\over 2},0\right)\oplus \left(0,{1\over 2}\right)\right]
$ with $K=1,3$, and $5$.
We first motivate legitimacy of such pattern as fundamental fields
as they emerge in the decomposition of triple fermion 
constructs into Lorentz representations.
We then study the baryon realization of RS fields as composite systems
by means of the quark version of the $U(4)$ symmetric
diatomic rovibron model. In using the 
$U(4)\supset O(4)\supset O(3)\supset O(2)$ 
reduction chain, we are able to reproduce quantum numbers 
and mass splittings of the above resonance assemblies. We present the
essentials of the four dimensional angular momentum algebra
and construct electromagnetic tensor operators. 
The predictive power of the model is illustrated by ratios of
reduced probabilities concerning electric de-excitations 
of various resonances to the nucleon. \\

\end{abstract}

\section{$O(4)$ Degeneracy Motif in Baryon Spectra: An Introduction}

One of the basic quality tests for any model of composite baryons is
the level of \hbox{accuracy} reached in describing the 
nucleon and $\Delta $ excitation spectra. 
In that respect, the knowledge on the degeneracy group of 
baryon spectra appears as a key tool in constructing the underlying 
Hamiltonian of the strong-interaction dynamics as a function
of the Casimir operators of the symmetry group.
To uncover the latter, one can analyze isospin by isospin how the 
masses of the resonances from the full baryon listing in Ref.~\cite{Part} 
spread with spin and parity. Such an analysis has been performed in 
prior work \cite{Ki97-98a} where it was found that 
Breit-Wigner masses reveal on the mass/spin ($M/J$) plane a 
well pronounced spin- and parity clustering. There, it was further shown 
that the quantum numbers of the resonances belonging to a particular 
cluster fit into O(1,3) Lorentz group representations of the so called
Rarita-Schwinger (RS) type \cite{RS}
\begin{equation}
\Psi_{\mu_1\mu_2...\mu_K}:=\left({K\over 2},{K\over 2}\right)\otimes
\left[\left({1\over 2},0\right)\oplus \left(0,{1\over 2}\right)\right]\, .
\label{RS_fields}
\end{equation}
To be specific, one finds the three RS clusters with $K=1,3$, and $5$ in
both the nucleon $(N)$ and $\Delta$ spectra. As long as the {\it Lorentz group
is locally isomorphic to $O(4)$\/}, multiplets with the quantum
numbers of the RS representations also appear in typical $O(4)$ problems such
as the levels of an electron with spin  in the hydrogen atom. 
There, the principal quantum number of the Coulomb problem is associated 
with $K+1$ while the r\'ole of the boost generators is taken by the 
components of the Runge-Lenz vector. 
The Rarita-Schwinger fields are the so-called ``diagonal case''
(i.e. $a=b={K\over 2}$) of the
more general representations $\left( a,b\right)\otimes
\left[\left({1\over 2},0\right)\oplus \left(0,{1\over 2}\right)\right]$.

\subsection{Rarita-Schwinger Fields as Multi--Spin-Parity States } 
The RS fields are described in terms of totally symmetric traceless
rank-$K$ Lorentz tensors with Dirac spinor components  
that satisfy the Dirac equation for each Lorentz index, $\mu_i$,
associated with a four-vector  $\left({1\over 2},{1\over 2}\right)$ 
space
\begin{eqnarray}	
\left(i\partial_\lambda \gamma^\lambda - M\right)\Psi_{\mu_1 \mu_2
\cdots \mu_K} = 0\, .
\label{Dirac_Proca}
\end{eqnarray}
The fields of the type in Eq.~(\ref{RS_fields}) were  considered six 
decades earlier by Rarita and Schwinger \cite{RS},
the most popular being the $K=1$-field
that has been frequently applied to the description of  spin-3/2 
particles.
Around mid sixties, Weinberg~\cite{We64} continued the 
tradition of the original Rarita-Schwinger work~\cite{RS} 
and considered  $\Psi_{\mu_1\mu_2...\mu_K}$ as fields suited for
the description of
{\it pure spin\/}-$J=K+{1\over 2}$ states of {\it fixed parity\/}.
The conjecture that $\Psi_{\mu_1\mu_2...\mu_K}$ can be reduced
to a single-spin state was based upon the belief that its 
lower-spin components are redundant, unphysical
states which can be removed  by means of the two auxiliary conditions 
$\partial^{\mu_1} \Psi_{\mu_1...\mu_K}=0$, and 
$\gamma^{\mu_1} \Psi_{\mu_1...\mu_K}=0$.
That these conditions do not serve the above purpose,
was demonstrated in Ref.~\cite{DVA_MK}.
There, the first auxiliary condition was shown to solely test 
consistency with the mass-shell relation $E^2-\vec{p}\, ^2=m^2$, 
while the second one amounted to the acausal energy-momentum dispersion
relation $E=-m\pm \sqrt{ \vec{p}\, ^2}$. It is that type of
acausality that must be at the heart of the Velo-Zwanziger problem \cite{VeZw}.
The RS fields in $O(4)$ are in fact compilations of fermions of different 
spins and parities. 
To illustrate this statement, and for the sake of concreteness, 
we here consider the coupling of,
say, a positive parity Dirac fermion to the 
$\left({K\over 2},{K\over 2}\right)$ hyper-boson
the latter being  composed of 
O(3) states of either natural $(\eta =+)$, or, 
unnatural $(\eta =- $) parities. 
These (mass degenerate) $O(3)$ states 
carry all integer internal angular momenta, $l$, with 
$l=0,\dots, K$ and transform (for the odd $K$'s of interest) with 
respect to the space inversion operation ${\cal P}$ according to
\begin{eqnarray}	
{\cal P}\vert K; \eta; lm\rangle  = \eta e^{i\pi l}
\vert K;\eta ; l\, -m\rangle \,, 
\quad
l^P=0^\eta,1^{-\eta},\dots, K^{-\eta}\, , \quad 
m=-l,\dots,l\, .
\label{parity}
\end{eqnarray}
In coupling now the Dirac spinor 
to $\left({K\over 2},{K\over 2}\right)$ from above, the following 
spin ($J$) and parity $(P$) quantum numbers  are created 
\begin{eqnarray}	
J^P = {1\over 2}^\eta,{1\over 2}^{-\eta}, {3\over 2}^{-\eta}, \dots,
\bigg(K+{1\over 2}\bigg)^{-\eta} \, .
\label{coupl_scheme}
\end{eqnarray}
In the following, we will use for  
the spin-sequence in Eq.~(\ref{coupl_scheme})
the short-hand notation $\sigma_ {2I, \eta}$, with $\sigma =K+1$, or, 
equivalently
\begin{equation}
\sigma_{2I, \eta }= \left( {{\sigma -1}\over 2}, {{\sigma -1}\over 2}\right)
\otimes \Big\lbrack \left( {1\over 2}, 0\right) \oplus 
\left( 0, {1\over 2}\right)\Big \rbrack\,
\chi^I .
\end{equation}
Here,  $\chi^I$ stands for the isospin spinor 
attributed to the states under consideration.

A glance at the baryon spectra teaches us that actually Nature strongly
favors the excitations of multi-spin-valued resonance clusters over 
that of pure higher-spin states.
This circumstance suggests a new data supported interpretation of the RS 
fields as complete resonance packages.

\subsection{Clustering Principle for Baryon Resonances}
In terms of the notations introduced above, all reported
light-quark baryons with masses \hbox{below} 2500~MeV (up to
the $\Delta \left(1600\right)$ resonance that is most probably an
independent quark-gluon hybrid state \cite{Barnes}), have been shown in 
Ref.~\cite{Ki97-98a} to be completely accommodated by the RS 
clusters $2_{2I,+} $, $4_{2I,-}$, and $6_{2I,-}$, having states 
of highest spin-$3/2^-$, $7/2^+$, and $11/2^+$, respectively 
(see Fig.~1). In each one of the nucleon, $\Delta$, and 
$\Lambda$ hyperon spectra, the natural parity cluster 
$2_{2I,+} $ is always of lowest mass. 
We consider it to reside in a Fock space, ${\cal F}_+$, 
built on top of a scalar vacuum. Equations (\ref{parity}) and 
(\ref{coupl_scheme}) illustrate how 
the $2_{2I,+}$ clusters (with  $I=1/2, 3/2$, and $0$) always unite the
first spin-${1\over 2}^+$, ${1\over 2}^-$, and ${3\over 2}^-$
resonances. {}For the non-strange baryons, $2_{2I,+}$
is followed by the unnatural parity clusters
$4_{2I,-}$, and $6_{2I,-}$, which we view to reside in a 
different Fock space, ${\cal F}_-$,  built on top of a 
pseudoscalar vacuum that is orthogonal (for an ideal $O(4)$ symmetry)
to the previous scalar vacuum.
To be specific, one finds all the seven 
$\Delta$-baryon resonances $S_{31}$, $P_{31}$, $P_{33}$, $D_{33}$, 
$D_{35}$, $F_{35}$ and $F_{37}$ from $4_{3,-}$ to be squeezed within 
the narrow mass region from 1900~MeV to 1950~MeV, while the $I=1/2$ 
resonances paralleling them, of which only the $F_{17}$ state is still
``missing'' from the data, are located around
1700$^{+20}_{-50}$~MeV (see left Fig.~1).

Therefore, the F$_{17}$ resonance 
is the only non-strange state with a mass below 2000 MeV which 
is ``missing'' for  the completeness of the present RS classification 
scheme. In further paralleling baryons from the third nucleon and
$\Delta$ clusters with $K+1=6$, one finds in addition the four
states $H_{1, 11}$, $P_{31}$, $P_{33}$, and $D_{33}$ with masses above
2000~MeV to be ``missing'' for the completeness of the new
classification scheme. The $H_{1, 11}$ state is needed to parallel the
well established $H_{3, 11}$ baryon, while the $\Delta$-states
$P_{31}$, $P_{33}$, and $D_{33}$ are required as partners to the (less
established) $P_{11}$(2100), $P_{13}$(1900), and $D_{13}$(2080)
nucleon resonances. 
{}For $\Lambda $ hyperons, incomplete data prevent a conclusive analysis.
Even so, Fig.~2 (left) indicates that 
the RS motif may already show up in the reported spectrum.  
The (approximate) degeneracy group of baryon spectra 
as already suggested in Refs.~\cite{Ki97-98a}, is, therefore, confirmed to be
\begin{equation}
SU(2)_I\otimes O(1,3)\simeq SU(2)_I\otimes O(4)\, ,
\label{isom}
\end{equation}
i.e., Isospin$\otimes$Space-Time symmetry.
To summarize, we here state the principle that
light unflavored baryon excitations are patterned after
Lorentz-multiplets. For example, 
the  Rarita--Schwinger spinors  $\Psi_{\mu_1...\mu_K}$ with
$K=1,3$, and $5$ accommodate all the $\pi N$ resonances
according to:
\begin{eqnarray}
{\cal F}_+:\quad 2_{2I ,+} \, :
\quad \Psi_{\mu_1} \, &:& P_{2I,1}; S_{2I,1}, 
D_{2I, 3}\, ,
\quad \mbox{for}\quad I= 0,\, {1\over 2},\, {3\over 2}\, ,
\quad \mbox{and}
\nonumber\\
{\cal F}_-:\quad 4_{2I, -} \, :\quad \Psi_{\mu_1\mu_2\mu_3}\, &:&
S_{2I,1};P_{2I,1}P_{2I,3};D_{2I,3},D_{2I,5}; F_{2I,5}, F_{2I,7}\, ,
\nonumber\\
{\cal F}_-:\quad 6_{2I, -} \, :\quad \Psi_{\mu_1\mu_2...\mu_5} \, &:&
S_{2I,1};P_{2I,1}P_{2I,3};D_{2I,3},D_{2I,5}; F_{2I,5}, F_{2I,7}\, ;\nonumber\\
&&G_{2I,7}, G_{2I,9};H_{2I,9},H_{2I,11}\, ,\quad
\mbox{for}\quad I= {1\over 2},\,  {3\over 2}\, ,\nonumber\\
\mbox{with the five ``missing'' states}:&\quad& 
F_{17}, H_{1,11}, P_{31}, P_{33}, D_{33}\, .
\label{class_scheme}
\end{eqnarray}
Occasionally, the above structures will be referred to as LAMPF clusters
to emphasize their close relationship to LAMPF physics.
\vspace{1truecm}
\begin{figure}[htb]
\vskip 5.0cm
\includegraphics{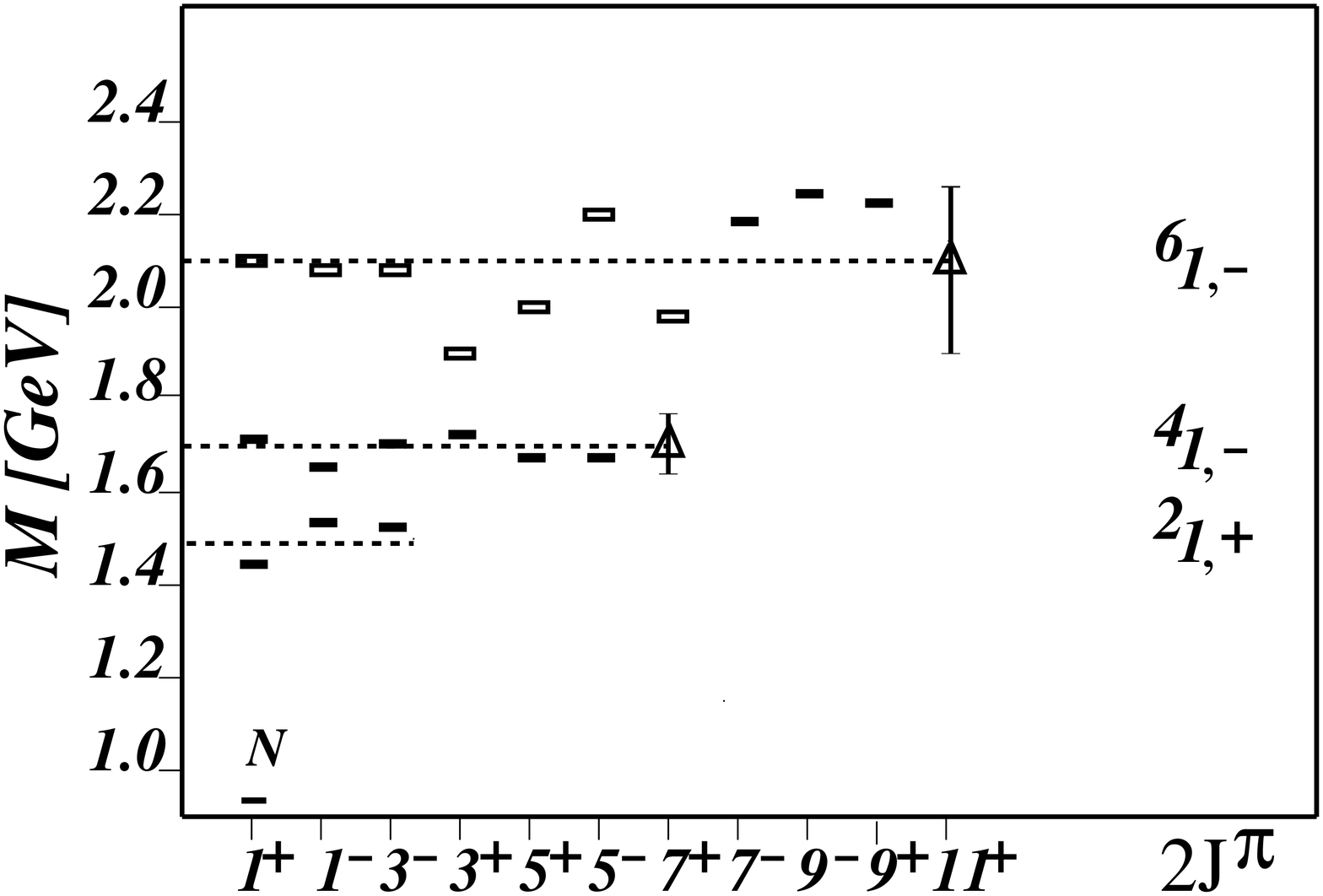}
\includegraphics{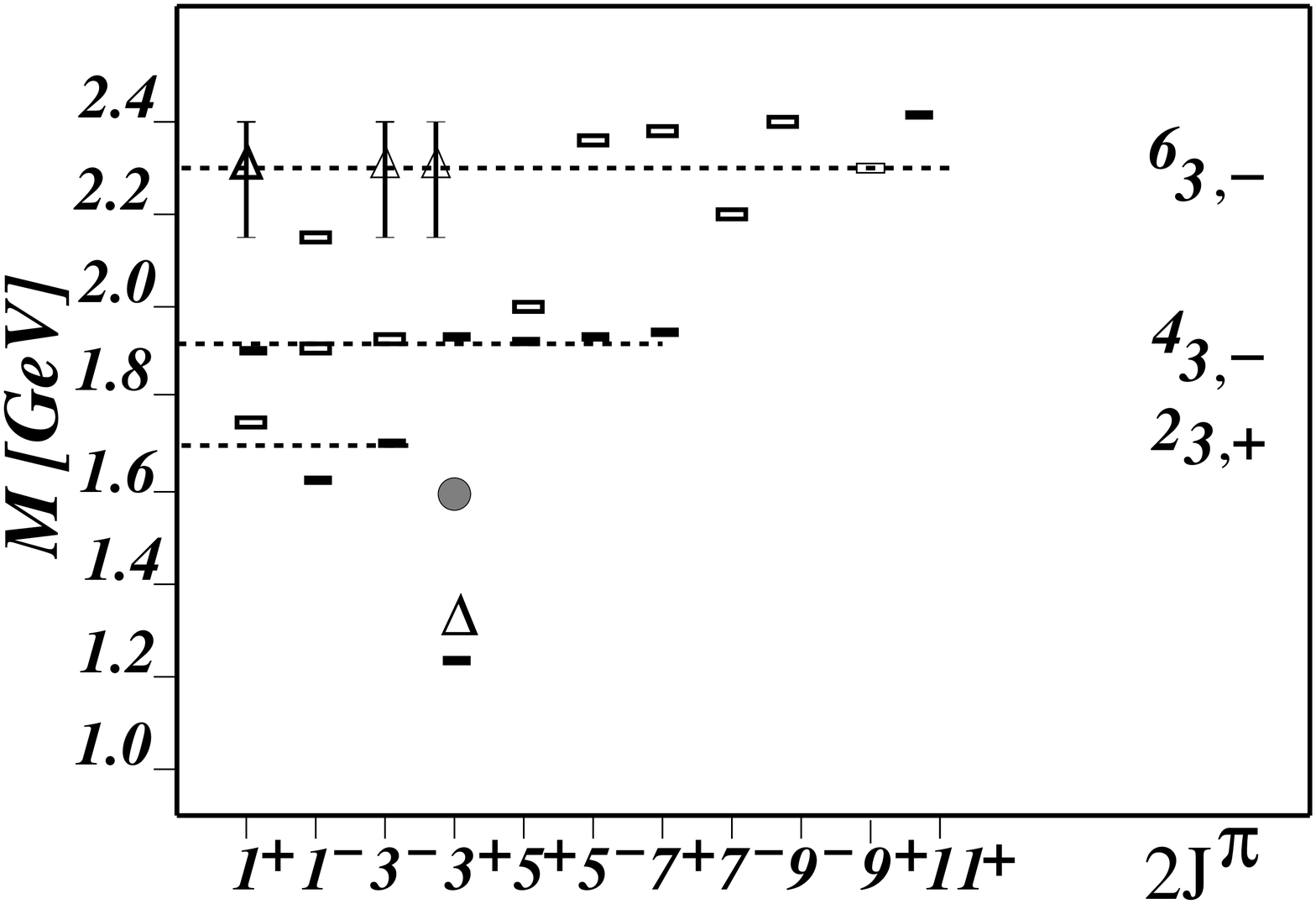}
\vspace{1.01cm}
{\small Fig.~1\hspace{0.2cm}
Rarita-Schwinger clustering of light unflavored baryon resonances.
The full bricks stand for three-to four-star resonances, the empty bricks
are one- to two-star states, while the triangles represent states that are
``missing'' for the completeness of the three RS clusters.
Note that ``missing''  $F_{17}$ and $H_{1,11}$ nucleon excitations 
(left figure) appear as  four-star resonances in the 
$\Delta $ spectrum (right figure). 
The ``missing'' $\Delta $ excitations $P_{31}$, $P_{33}$, and $D_{33}$
from $6_{3,-}$ are one-to two star resonances in the nucleon counterpart
$6_{1,-}$. The $\Delta (1600)$ resonance (shadowed oval) drops 
out of our RS cluster systematics and we view it
as an independent hybrid state.}
\label{fig:RS_degeneracy}
\end{figure}
The scalar vacuum in the first Fock space reflects the Nambu-Goldstone
mode of chiral symmetry near the ground state.
As argued in Ref.~\cite{MK2000}, its change to a pseudoscalar
between the 1st and 2nd clusters, may  be related to a change of 
the mode of chiral symmetry realization in baryonic spectra.

\vspace*{0.21truein}

\begin{figure}[htb]
\vskip 5.0cm
\includegraphics{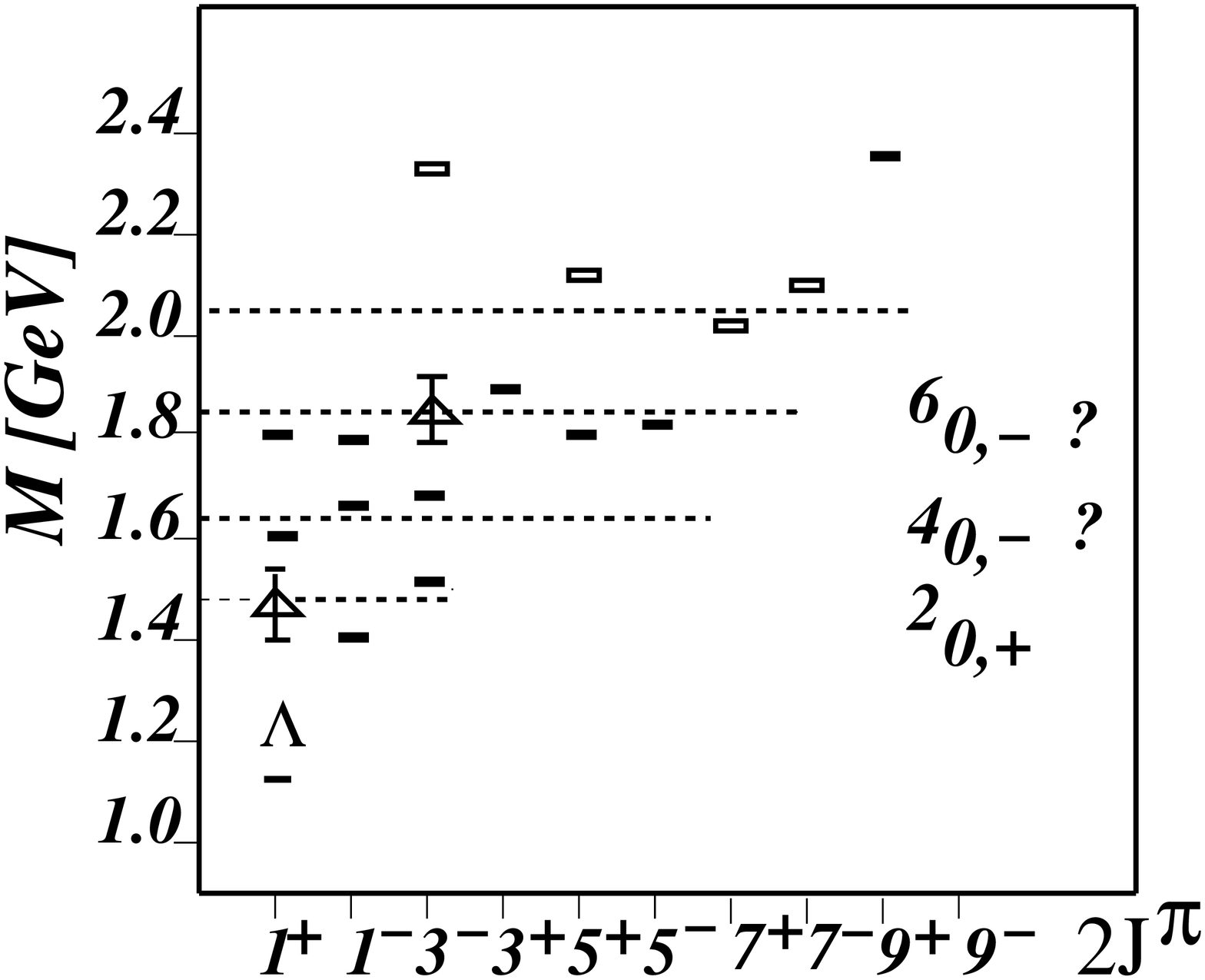}
\includegraphics{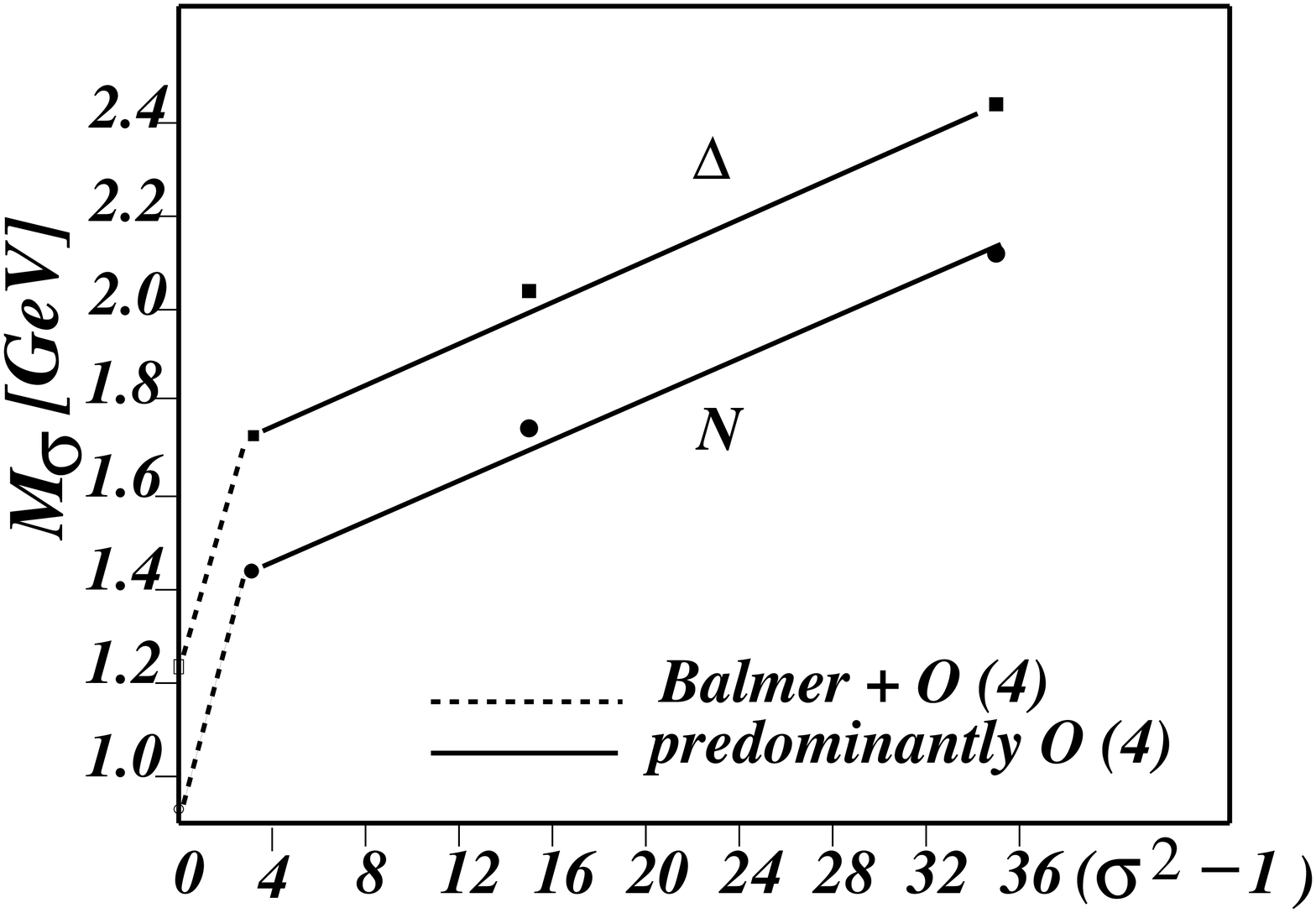}
\vspace{1.01cm}
{\small Fig.~2\hspace{0.2cm}
Clustering traces in the $\Lambda $ hyperon spectrum (left). 
O(4) rotational bands of nucleon (N) and $(\Delta )$ excitations
(right). Notations as in Fig.~1.} 
\label{fig:O4_band}
\end{figure}

Within our scheme, the inter-cluster spacing of $200$ to $300$~MeV 
is larger by a factor of $3$ to $6$ as compared
to the mass spread within the clusters. For example, the $2_{1,+}$,
$2_{3,+}$, $4_{1,-}$, and $4_{3,-}$ clusters carry
the maximal internal mass splitting of $50$ to $70$~MeV.

Finally, the reported mass averages of the resonances from the 
RS multiplets with $K=1,3$, and $5$ are well described by means of the
following simple {\it empirical\/} relation:
\begin{eqnarray}	
M_{{\sigma};I } =  M_I -m_1\, {1\over {\sigma^2}}   +  m_2 \, 
{{\sigma ^2-1}\over 4}\, , \quad I={1\over 2},\, \, {3\over 2}\,  , 
\label{Balmer_ser}
\end{eqnarray}
where, again, $\sigma=K+1$.
The parameters take for the nucleon $(I={1\over 2}$)
the values $m_1=600$~MeV, 
$m_2=70$~MeV, and $M_{1\over 2}=M_N+m_1$,
respectively. The $\Delta $ spectrum ($I={3\over 2})$
is best fitted by the smaller $m_2$ value of $m_2=40$ MeV and 
$M_{3\over 2}=M_\Delta +m_1$ (right Fig.~2 ).

It is the goal of this paper to develop a constituent model for
baryons that explains the observed clustering in the spectra
of the light unflavored baryons. The paper is organized as follows.
In Section II we motivate legitimacy of  
fundamental fields of specified mass and 
unspecified spin as they emerge in the decomposition of a 
triple-Dirac-fermion
system into Lorentz group representations.
In Section III we present the quark version of the diatomic rovibron 
model \cite{Diat} and study its excitation modes. There we also establish
correspondence between excited rovibron states and the  baryonic
RS clusters. We further make all the observed and some of the 
``missing'' resonances distinguishable in organizing them into different
rovibron modes. We construct the relevant quark Hamiltonian and 
recover Eq.~(\ref{Balmer_ser}). We finally outline the construction
of electric transition operators and
calculate selected electric transitions of cluster inhabitants
to the nucleon.
The paper is finished by a brief summary and outlook.

\section{Multi-Spin States As Lorentz Covariant Representations}

The relativistic description of three-Dirac-spinor systems 
has been studied in detail in
Ref.~\cite{MoNiShaSmi}. Starting with the well known Lorentz 
invariance of the ordinary Dirac equation
\begin{equation}
(\gamma^\mu p_\mu -m)\, u(\vec{p}\, ) =0\, ,
\label{Dir_eq}
\end{equation}
the authors show that the direct product of three Dirac spinors
gives rise to a 64-dimensional linear equation of the type
\begin{eqnarray}
(\Gamma^\mu p_\mu -m)\, {\cal U}(\vec{p}\, ) =0\, ,\quad &\mbox{with}&\quad
\Gamma^\mu =\sum_{r=1}^3 \gamma_r^\mu \, \nonumber\\
\gamma_1^\mu =\gamma^\mu \otimes I\otimes I,
\quad 
\gamma_2^\mu =I\otimes\gamma^\mu\otimes I, &\quad &
\gamma_3^\mu =I\otimes I\otimes \gamma^\mu,
\label{Bhabha}
\end{eqnarray}
Here, $I$ stays for the four dimensional unit matrix, while
the index $r$ indicates position of 
the Dirac matrix $\gamma^\mu$ in $\gamma_r^\mu $.
Under Lorentz transformations ($a^\mu_\nu$) of the $\gamma $ matrices, 
the matrices $\Gamma^\mu $ from Eq.~(\ref{Bhabha})
change according to $\Gamma^\mu$' =$U\Gamma^\mu U^{-1}$ with 
$U=U_1\otimes U_2\otimes  U_3$, and  
$ U_r$ defined as the matrix
that covers the Lorentz transformation
 $\gamma^\mu_r$'=$a^\mu_\nu \gamma^\nu_r$
 = $ U_r\gamma^\mu_r  U_r^{-1}$ of  $\gamma_r $. 
Equation (\ref{Bhabha}) is therefore Lorentz 
invariant. Moreover, it was demonstrated that Eq.~(\ref{Bhabha})
has $U(4)$ as an additional dynamical symmetry.

The 64 states from above are distributed over different irreducible 
representations (irreps) of $U(4)$ and the permutational group
group ${\cal S}_3$ as well. To be specific, one finds two  20plets
in turn associated with the Young schemes $\left[ 3000\right]$,
and $\left[ 2100 \right]$. They are completed by the quartet
$\left[ 1110 \right]$. The three-Dirac spinor state (denoted by $s^3$)
can be characterized by the set of quantum numbers
\begin{equation}
\vert s^3 \lbrack f\rbrack X, \lbrace f\rbrace R\rangle\, .
\label{q3_64}
\end{equation}
Here, $X$ stands for a set of quantum numbers characterizing the $U(4)$ 
basis vectors of the $\lbrack f\rbrack $ irrep, while $R$ denotes the 
Yamanouchi symbol
labeling the basis vectors of the ${\cal S}_3$ representation 
$\lbrace f\rbrace $ \cite{Wy}. The Yamanouchi symbols for the 
$\lbrack 3000 \rbrack$,  $\lbrack 2100\rbrack $, and
$\lbrack 1110 \rbrack $ are 
1; 2, 1,  and 1, respectively.
The complete number $(N_{s^3}$) of 64 states of the three-Dirac-fermion
($s^3$) system  is then encoded 
by the relation
\begin{equation}
N_{s^3 }=\sum_{\lbrack f\rbrack } \mbox{dim} \, \lbrack f \rbrack \, 
\mbox{dim}\,\lbrace f\rbrace\, 
\label{64} 
\end{equation}
where dim$\lbrack f\rbrack$ and dim$\lbrace f\rbrace$ are in turn the 
dimensionalities
of the $U(4)$ irrep $\lbrack f\rbrack$, and the ${\cal S}_3$ irrep 
$\lbrace f\rbrace$, respectively.
In considering now the reduction chain $U(4)\supset O(5)$, allows for a more 
detailed specification
of the spin content of the $U(4)$ multiplets from above 
(see Ref.~\cite{MoNiShaSmi} for details).

The quantum numbers of the irreducible representations
(irreps) of $O(5)$ are labeled by the two numbers 
$(\lambda_1\, \lambda_2)$ which can be either integer, 
or half-integer.
The states participating a given
$O(5)$ irrep can be further specified by 
the quantum numbers of the irreps of the $O(5)$ subgroups appearing
in the reduction chain $O(5)\supset O(4) \supset O(3)\supset O(2)$. 
To specify  the $O(4)$ irreps in the context of the $O(5)$ reduction
down to $O(2)$ it is more convenient to use instead of the
pair $\left(a, b\right)$ from above, rather the pair 
$\left(m_1\, m_2\right )$ with the mapping 
\begin{equation}
m_1= a+b \, , \qquad m_2=a-b \, .
\label{mapping_jj_mm}
\end{equation}
{}Finally, the $O(3)$ irreps in the $O(3)\supset O(2) $ reduction scheme
are labeled by the well known spin (J) and magnetic
quantum number (M). The complete set of quantum numbers specifying a
member of a $O(5)$ multiplet 
$ \vert \left(\lambda_1\, \lambda_2\right);\left(m_1\, m_2\right);JM\rangle $
satisfy the inequalities
\begin{eqnarray}
\lambda_1\ge m_1\ge\lambda_2\ge \vert m_2\vert  \,  ,\nonumber\\
m_1\ge J\ge |m_2|\, , &\quad& J\ge M\ge -J \, .
\label{Schachtelung}
\end{eqnarray}
The $U(4)$ irrep $\lbrack 2100\rbrack $ is of particular interest for the 
present work. In the $U(4)\supset O(5)$ reduction chain it splits into 
$O(5)$ irreps according to
\begin{equation}
\lbrack 2100\rbrack\longrightarrow \left({3\over 2}{1\over 2}\right)
\oplus\left( {1\over 2}{1\over 2}\right)\, .
\end{equation}
The first irrep on the rhs of the last equation is 16-dimensional, 
while the second is four dimensional and associated with a Dirac spinor.
As we shall see below, the  $O(5)$ 16plet 
$\left({3\over 2}\, {1\over 2}\right)$
is nothing but the RS field with $ K=1$. Indeed,  
from Eq.~(\ref{Schachtelung}) follows that
\begin{equation}
{3\over 2}\ge m_1\, , \quad m_1 \ge {1\over 2}\, ,
\quad \mbox{and} \quad {1\over 2} \ge |m_2|\, .
\label{RS_O5}
\end{equation}
The inequalities in the latter equation are satisfied
for $m_1=3/2$, $1/2$, and for $m_2=1/2$, $-1/2$.
In accordance with the 2nd equation in (\ref{Schachtelung}),
$J$ can take the three values $J=3/2, 1/ 2$, and $J=1/2$.  
Thus, the $\left({3\over 2}{1\over 2}\right)$ irrep of O(5)
describes a spin-3/2 and two spin-1/2 states and
coincides with the lowest 16-dimensional Rarita-Schwinger field.

The above consideration gives an idea of how Lorentz representations
of the RS type can emerge as {\it fundamental\/}
free particles of definite mass and indefinite spin within the context 
of a relativistic space-time treatment.
Though such point-like particles have not been detected so far,
the $N$ and $\Delta $ spectra strongly indicate existence of 
{\it composite\/} RS fields.
In the following, we shall focus onto that very realization of multi-spin  
Lorentz representations and explore their internal structure by means of 
constituent models.
{}For a more profound textbook presentation on the
various aspects of higher-dimensional relativistic supermultiplets,  
the interested reader is referred to 
Ref.~\cite{MoSmi}.

\section{The Quark Version of the Diatomic Rovibron Model and the
RS Clustering in Baryon Spectra}
 
Baryons in the quark model are considered as constituted of three
quarks in a color singlet state. It appears naturally, therefore,
to undertake an attempt of describing the baryonic system
by means of algebraic models developed for
the purposes of triatomic molecules,
a path already pursued by  Refs.~\cite{doublets}.
There, the three body system was described in terms of two
vectorial ($\vec{p}\, ^+$) and one scalar $(s^+)$ boson degrees of freedom
that transform as the fundamental  $U(7)$ septet.
In the  dynamical symmetry limit 
\begin{equation}
U(7)\longrightarrow U(3)\times U(4)
\label{U(7)}
\end{equation}
the degrees of freedom associated with the one vectorial boson
factorize from  those associated with
the scalar boson and the remaining vectorial boson.
Because of that the physical states constructed within the $U(7)$ IBM model
are often labeled by means of  $U(3)\times U(4)$ quantum numbers.
Below we will focus on that very sub-model of the IBM and
show that it perfectly accommodates the RS clusters from above
and thereby the LAMPF data on the non-strange baryon
resonances.  

The dynamical limit $U(7
)\longrightarrow U(3)\times U(4)$ corresponds
to the quark--diquark approximation of the three quark system, when two
of the quarks reveal a stronger pair correlation to a diquark (Dq)
~\cite{Anselmino}, while the third quark (q) acts as a spectator.
The diquark approximation turned out to be rather convenient
in particular in describing various properties of the ground state 
baryons~\cite{Hel97,Kus}. Within the context of the quark--diquark (q-Dq) 
model, the ideas of the rovibron model, known from the spectroscopy of 
diatomic molecules ~\cite{Diat}, can be applied to the description
of the rotational-vibrational (rovibron) excitations of the q--Dq system.

\subsection{Rovibron Model for the Quark--Diquark System}

In the rovibron model (RVM) the relative q--Dq motion (see Fig.~3)
is described by means of four types of boson creation
operators $s^+, p^+_1, p^+_0$, and $p^+_{-1}$ 
(compare ~\cite{Diat}). The operators $s^+$ and $p^+_m$ in turn 
transform as rank-0, and rank-1 spherical tensors,
i.e. the magnetic quantum number $m$ takes
in turn the values $m=1$, $0$, and $-1$.
In order to construct boson-annihilation operators that
also transform as spherical tensors, one introduces
the four operators $\tilde{s}=s$, and
$\tilde{p}_m=(-1)^m\, p_{-m}$.
Constructing rank-$k$ tensor product of
any rank-$k_1$ and rank-$k_2$ tensors, say, $A^{k_1}_{m_1}$ 
and $A^{k_2}_{m_2}$, is standard and given by
\begin{equation}
\lbrack A^{k_1}\otimes A^{k_2}\rbrack^k_m =
\sum_{m1,m_2}(k_1m_1 k_2m_2\vert km)\, A^{k_1}_{m_1}A^{k_2}_{m_2}\, .
\label{clebsh}
\end{equation}
Here, $(k_1m_1k_2m_2\vert km)$ are the well known $O(3)$ Clebsch-Gordan
coefficients.
\begin{figure}[htbp]
\centerline{\psfig{figure=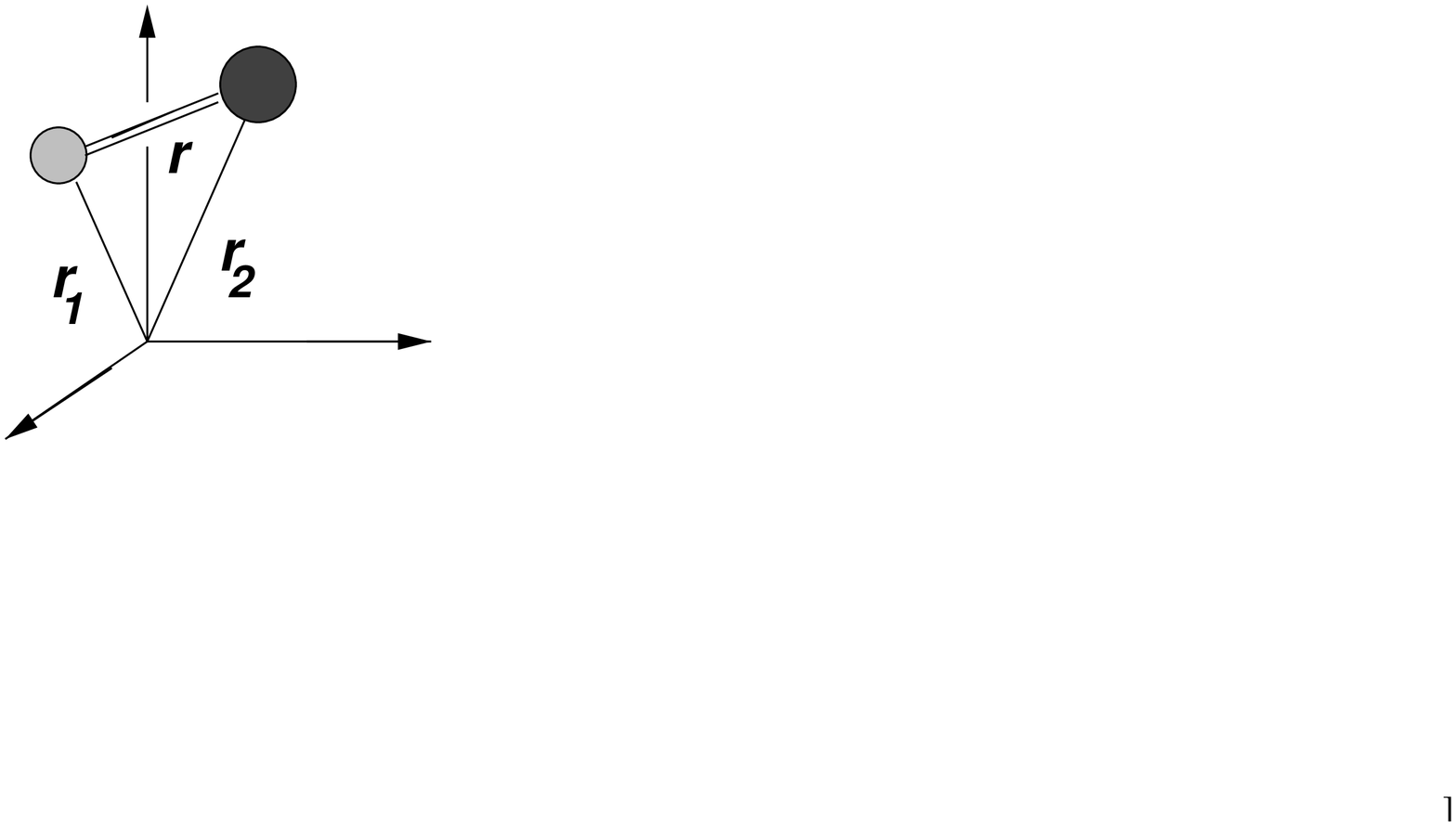,width=10cm}}
\vspace{0.1cm}
{\small Fig.~3\hspace{0.2cm} 
Schematic presentation of a q-Dq two-body system.
}
\label{fig:ro_vibron}
\end{figure}

Now, the lowest states of the two-body system are identified with $N $
boson states and are characterized by the ket-vectors 
$\vert n_s\, n_p\, l\, m\rangle $ (or, a linear combination of them)
within a properly defined Fock space. The constant  
$N=n_s +n_p$ stands for the total number of $s$- and $p$ bosons
and plays the r\'ole  of a parameter of the theory.
In molecular physics, the parameter $N$ is usually associated
with the number of molecular bound states.
The group symmetry of the rovibron model is well known to be $U(4)$. 
The fifteen generators of the associated $su(4)$ algebra 
are determined as the following set of bilinears 
\begin{eqnarray}
A_{00}=s^+ \tilde{s}\, , &\quad& A_{0m}=s^\dagger \tilde{p}_m\, , 
\nonumber\\
A_{m0}=p^+_m\tilde{s}\, , &\quad & A_{mm'}=p^\dagger _m\tilde{p}_{m'}\, .
\label{RVM_u4}
\end{eqnarray} 
The $u(4)$ algebra is then recovered by the following 
commutation relations
\begin{equation}
\lbrack A_{\alpha\beta},A_{\gamma\delta}\rbrack_-=
\delta_{\beta \gamma}A_{\alpha\delta}-
\delta_{\alpha\delta}A_{\gamma\beta}\, .
\end{equation}
The operators associated with physical observables can then be expressed
as combinations of the $u(4)$ generators.
To be specific, the three-dimensional angular momentum takes the form
\begin{equation}
L_m=\sqrt{2}\, \lbrack p^+ \otimes \tilde{p}\rbrack^1_m \, .
\label{a_m}
\end{equation}
Further operators are  $(D_m$)-- and $(D'_m$) defined as 
\begin{eqnarray}
D_m &=&\lbrack p^+\otimes \tilde{s}+s^+\otimes \tilde{p}\rbrack^1_m\, ,\\
\label{x_dipol_rvm}
D_m '&=&i\lbrack p^+\otimes \tilde{s}-s^+\otimes \tilde{p}\rbrack^1_m\, ,
\label{p_dipol_rvm}
\end{eqnarray}
respectively.
Here, $\vec{D}\, $ plays the r\'ole of
the electric dipole operator.

{}Finally, a quadrupole operator $Q_m$ can be constructed as
\begin{equation}
Q_m=\lbrack p^+\otimes \tilde{p}\rbrack^2_m\, ,
\quad \mbox{with}\quad m=-2,..., +2\, .
\label{quadr_rvm}
\end{equation}
The $u(4)$ algebra has the two algebras $su(3)$, and $so(4)$, as respective
sub-algebras. The $su(3)$ algebra is constituted by the three
generators $L_m$, and the five components of the quadrupole operator
$Q_m$. Its $so(4)$ subalgebra is constituted by the 
three components of the angular momentum operator $L_m$, on the one side,
and the three components of the operator $D_m'\, $, on the other side. 
Thus there are two exactly soluble RVM limits that correspond 
to the two different chains of reducing
$U(4)$ down to $O(3)$. These are:
\begin{equation}
 U(4)\supset U(3)\supset O(3)\, \quad \mbox{and} \quad  
U(4)\supset O(4)\supset O(3)\, ,
\label{chains}
\end{equation}
respectively.
The Hamiltonian of the RVM in these exactly soluble limits is 
then constructed as a properly chosen function of the Casimir 
operators of the algebras of either the first, or the second chain. 
{}For example, in case one approaches $O(3)$ via $U(3)$,
the Hamiltonian of a dynamical $SU(3)$ symmetry can be cast into 
the form:
\begin{equation}
H_{SU(3)} = H_0 +\alpha\,  {\cal C}_2\left( SU(3) \right) +
\beta \, {\cal C}_2\left( SO(3) \right)\, .
\label{H_SU3}
\end{equation}
Here, $H_0$ is a constant,
${\cal C}_2\left( SU(3) \right)$, and ${\cal C}_2\left( SO(3) \right)$ are 
in turn the quadratic (in terms of the generators) 
Casimirs of the groups $SU(3)$, and $SO(3)$, 
respectively, while $\alpha $ and $\beta $ are constants, to
be determined from data fits.

A similar expression (in obvious notations) can be written for the 
RVM Hamiltonian in the $U(4)\supset O(4)\supset O(3)$ exactly soluble limit:
\begin{equation}
H_{SO(4)} = H_0 +\widetilde{\alpha}\, {\cal C}_2\left(SO(4)\right) +
\widetilde{\beta} {\cal C}_2\left(SO(3)\right)\, .
\label{H_SO4}
\end{equation}
The Casimir operator ${\cal C}_2\left(SO(4)\right)$ is defined accordingly as
\begin{equation}
{\cal C}_2\left( SO(4)\right)={1\over 4}\left( \vec{L}\, ^2 + 
\vec{D}\, ' \, ^2
\right)\, 
\label{so(4)_Casimir}
\end{equation}
and has an eigenvalue of ${K\over 2}\left( {K\over 2}+1 \right)$.
In molecular physics, only linear combinations of the Casimir operators 
are used, as a rule. However, as known
from the hydrogen atom \cite{ElDo}, the Hamiltonian is determined by the 
inverse  power of ${\cal C}_2\left(SO(4)\right) $ according to
\begin{equation}
H_{Coul}=f \left( -4C_2\left(SO(4)\right) -1\right)^{-1}
\label{Coul+SO(4)}
\end{equation}
where $f$ is a parameter with the dimensionality of mass.
This Hamiltonian predicts the energy of the states as $E_K=-f/(K+1)^2$ 
and does not follow the simple linear pattern
(see also Eq.~(\ref{H_SO4})).

In order to demonstrate how the RVM applies to baryon spectroscopy,
let us consider the case of q-Dq states associated with $N=5$
and for the case of a $SO(4)$ dynamical symmetry. 
{}From now on we shall refer to the quark rovibron model
as qRVM. It is of common
knowledge that the totally symmetric irreps of the $u(4)$ algebra 
with the Young scheme $\lbrack N\rbrack$ contain the 
$SO(4)$ irreps $\left({K\over 2}, {K\over 2}\right)$ with
\begin{equation}
K=N, N-2, ..., 1 \quad \mbox{or}\quad 0\, .
\label{Sprung_K}
\end{equation}
Each one of these $SO(4)$ irreps contains $SO(3)$ multiplets with
three dimensional angular momenta
\begin{equation}
l=K, K-1, K-2, ..., 1, 0\, .
\label{O3_states}
\end{equation}
In applying the branching rules in Eqs.~(\ref{Sprung_K}),
(\ref{O3_states})
to the case $N=5$, one encounters the series of levels
\begin{eqnarray}
K&=&1: \quad l=0,1;\nonumber\\
K&=&3: \quad l=0,1,2,3;\nonumber\\
K&=&5: \quad l=0,1,2,3,4,5\, .
\label{Ns_Ks}
\end{eqnarray}
The parity carried by these levels is $\eta (-1)^{l}$ where
$\eta $ is the parity of the relevant vacuum. In coupling now the
angular momenta in Eq.~(\ref{Ns_Ks}) to the spin-1/2 of the three
quarks in the nucleon, the following sequence of states is obtained:
\begin{eqnarray}
K&=&1: \quad \eta J^\pi={1\over 2}^+,{1\over 2}^-, {3\over 2}^-\, ;
\nonumber\\
K&=&3: 
\quad \eta J^\pi={1\over 2}^+,{1\over 2}^-, {3\over 2}^-,
{3\over 2}^+, {5\over 2}^+, {5\over 2}^-, {7\over 2}^- \, ;
\nonumber\\
K&=&5: \quad \eta J^\pi={1\over 2}^+,{1\over 2}^-, {3\over 2}^-,
{3\over 2}^+, {5\over 2}^+, {5\over 2}^-, {7\over 2}^- ,
{7\over 2}^+, {9\over 2}^-, {11\over 2}^-\, .
\label{set_q}
\end{eqnarray}
Thus rovibron states of half-integer spin will transform according to  
$\left( {K\over 2},{K\over 2}\right) \otimes \left[
\left({1\over 2},0 \right) \oplus 
\left( 0,{1\over 2} \right)\right] $
representations of $SO(4)$.
The isospin structure is accounted for pragmatically through
attaching to the RS clusters an isospin spinor
$\chi^I$ with $I$ taking the values $I={1\over 2}$ 
and $I={3\over 2}$ for the nucleon, and the $\Delta $ states,
respectively.
As illustrated by Fig.~1, the above quantum numbers cover
both the nucleon and the $\Delta $ excitations.

Note that in the present simple version of the
rovibron model, the spin of the quark--diquark system is
$S={1\over 2}$, and the total spin $J$ takes the values $J=l\pm {1\over 2}$
in accordance with Eqs.~(\ref{Ns_Ks}) and (\ref{set_q}).
The strong relevance of {\it same \/} picture for both
the nucleon and the $\Delta (1232) $ spectra (where
the diquark is in a vector-isovector state) hints onto the
dominance of a scalar  diquark for both the excited nucleon-- and  
$\Delta (1232)$ states.
This situation is reminiscent of the $^210$ configuration
of the $70(1^-)$plet of the canonical 
$SU(6)_{SF}\otimes O(3)_L$ symmetry
where the mixed symmetric/antisymmetric character
of the $S=1/2$ wave function in spin-space is compensated
by a mixed symmetric/antisymmetric wave function in coordinate space,
while the isotriplet $I=3/2$ part is totally symmetric.

We here will leave aside the discussion of the generic problem
of the various incarnations of the IBM model regarding the
symmetry properties of the resonance wave functions to a later
date and rather concentrate in the next subsection onto 
the ``missing'' resonance problem.

\subsection{Observed and ``Missing'' Resonance Clusters within the 
Rovibron Model}

The comparison of the states in Eq.~(\ref{set_q}) with the reported ones
in  Eq.~(\ref{class_scheme}) shows that the predicted sets are in agreement 
with the characteristics of the non-strange baryon excitations with
masses below $\sim $ 2500 MeV, provided, the parity $\eta $ of the vacuum
changes from scalar ($\eta =1$) for the $K=1$, to pseudoscalar 
($\eta =-1$) for the $K=3,5$ clusters.
A pseudoscalar ``vacuum'' can be modeled in terms of
an excited  composite  diquark carrying an internal 
angular momentum  $L=1^-$ and maximal spin $S=1$. In one of
the possibilities the total spin of such
a system can be $\vert L-S\vert = 0^-$.
To explain the properties of the ground state, one has to
consider separately even $N$ values, such as, say, $N'=4$.
In that case another branch of excitations, with $K=4$, $2$, and
$0$ will emerge. The $K=0$ value characterizes the 
ground state,  $K=2$ corresponds to
$\left( 1,1\right)\otimes 
\lbrack\left( {1\over 2},0\right)\oplus 
\left(0,{1\over 2}\right) \rbrack $, while  $K=4$ corresponds to
$\left( 2,2\right) \otimes\lbrack\left( {1\over 2},0\right)\oplus 
\left(0,{1\over 2}\right) \rbrack $. 
These are the multiplets that we 
will associate with the  ``missing'' resonances  
predicted by the rovibron model.
In this manner, reported and ``missing'' resonances  fall apart 
and populate  distinct $U(4)$- and $SO(4)$ representations.
In making observed and ``missing'' resonances distinguishable,
reasons for their absence or, presence in the spectra are
easier to be searched for.
As to the parity of the resonances with even $K$'s, there is some ambiguity.
As a guidance one may consider the decomposition of the three-quark
($q^3$) Hilbert space into Lorentz group representations
as performed in Ref.~\cite{MK2000}. There, two states of 
the type
$\left( 1,1\right)\otimes 
\lbrack\left( {1\over 2},0\right)\oplus 
\left(0,{1\over 2}\right) \rbrack $
were found. The first one arose out of the decomposition of 
the $q^3$-Hilbert space spanned by the $1s-1p-2s$ single-particle
states. It was close to  $\left( {1\over 2},{1\over 2}\right)\otimes 
\lbrack\left( {1\over 2},0\right)\oplus 
\left(0,{1\over 2}\right) \rbrack $
and carried opposite parity to the latter.
It accommodated, therefore, 
unnatural parity resonances. 
The second $K=2$ state was part of the
$(1s-3s-2p-1d)$- single-particle configuration space and
was closer to $\left( {3\over 2},{3\over 2}\right)\otimes 
\lbrack\left( {1\over 2},0\right)\oplus 
\left(0,{1\over 2}\right) \rbrack $.
It also carried opposite parity to the latter and accommodated
natural parity resonances. Finally, the $K=4$ cluster 
$\left( 2,2\right)\otimes 
\lbrack\left( {1\over 2},0\right)\oplus 
\left(0,{1\over 2}\right) \rbrack $
emerged in the decomposition of the one-particle-one-hole states within
the $(1s-4s-3p-2d-1f-1g)$ configuration space and
carried also natural parity, that is, opposite parity to
  $\left( {5\over 2},{5\over 2}\right)\otimes 
\lbrack\left( {1\over 2},0\right)\oplus 
\left(0,{1\over 2}\right) \rbrack $.
In accordance with the above results,
we here will treat the $N=4$ states to be all of natural parities
and identify them with the nucleon $(K=0)$, the natural parity $K=2$,
and  the natural parity $K=4$ RS clusters.

The unnatural parity $K=2$ cluster from \cite{MK2000}
could be generated through an unnatural
parity $N=2$ excitation mode. 
However, this mode would require manifest chiral symmetry up to 
$\approx 1550$
MeV which contradicts at least present data.
With this observation in mind, we here will restrict ourselves
to the consideration of the natural parity $N=4$ clusters.
In this manner the unnatural parity $K=2$ state from Ref.~\cite{MK2000}
will be dropped out from the current version of the rovibron model.
{}From now on we will refer to the excited $N=4$ states as to
``missing'' rovibron  clusters.

Now, the qRVM Hamiltonian that reproduces the mass
formula from Eq.~(\ref{Balmer_ser}) is given by the following
function of ${\cal C}_2\left(SO(4)\right)$
\begin{equation}
H_{qRVM}=H_0 - f_1\, \left(4 {\cal C}_2\left( SO(4)\right) +1\right)^{-1}
+f_2\, \left( {\cal C}_2(SO(4)\right)\, .
\label{H_QRVM}
\end{equation}
The states in  Eq.~(\ref{set_q}) are degenerate and the dynamical symmetry
is $SO(4)$. The parameter set 
\begin{equation}
H_0= M_{N/\Delta } +f_1\, ,\quad  f_1=m_1\, , \quad f_2=m_2\, ,
\label{f_s}
\end{equation}
with  $I={1\over 2},{3\over 2} $, recovers the empirical mass formula in 
Eq.~(\ref{Balmer_ser}).
Thus, the $SO(4)$ dynamical symmetry limit of the
qRVM picture of baryon structure motivates  
existence of quasi-degenerate clusters of resonances 
in the nucleon- and $\Delta $ baryon spectra.
In Table I we list the masses of the RS clusters concluded from
Eqs.~(\ref{H_QRVM}), and (\ref{f_s}).

\begin{table}[htbp]
\caption{
Predicted mass distribution of observed (obs), and
missing (miss) rovibron  clusters (in MeV) according to
Eq.~(34,35). The sign of $\eta $ in Eq.~(3) determines natural-
($\eta =+1$), or, unnatural ( $\eta =-1$) parity states.
{}All $\Delta $ excitations have been calculated
with  $m_2=40$ MeV rather than
with the nucleon value of $m_2=70$ MeV. 
The experimental mass averages of the resonances from a given
RS cluster have been labeled by ``exp''.
The nucleon and $\Delta $ ground state masses  
$M_N$ and $M_\Delta $ were taken to equal their experimental values. }

\vspace*{0.21truein}
\begin{tabular}{lccccccc}
\hline
~\\
K & sign $\eta $ & N$^{\mbox{obs} }$  & N$^{\mbox{exp}}$ &
 $ \Delta^{\mbox{obs}}$ & $\Delta^{\mbox{exp}}$ &  
   N$ ^{\mbox{miss}}$ & $\Delta^{\mbox{miss}}$  \\
\hline
~\\
0 & + &939 &939 & 1232& 1232 & &  \\
1&+  &1441 & 1498 & 1712 &1690 &         & \\
2&+  &     &      &      &     &  1612   & 1846 \\
3&-  &1764 &1689  & 1944 &1922 &         &     \\
4&+  &     &      &      &     &  1935   & 2048 \\
5&-  &2135 &2102  & 2165 &2276 &         &      \\
\hline
\end{tabular}
\end{table}

The data on the $\Lambda $, $\Sigma $, and $\Omega^-$ hyperon spectra
are still far from being as complete as those of the nucleon and the 
$\Delta $ baryons and do not allow, at least at the present stage, 
a conclusive statement on relevance or irrelevance
of the rovibron picture Fig.~1.   
The presence of the heavier strange quark can significantly 
influence the excitation modes of the $q^3$-system.
In case, the presence of the $s$ quark in the
hyperon structure is essential,  the
$U(4)\supset U(3)\supset O(3)$ chain can be favored
over $U(4)\supset O(4)\supset O(3)$ and a different clustering motif
can appear here. For the time being, this issue
will be dropped out of further consideration.

In the next subsection, we shall outline the calculational
scheme for branching ratios of reduced probabilities for
electromagnetic transitions.

\subsection{O(4) Angular Momentum Algebra and Multipole Operators}

In the following, resonance states from a RS cluster will be denoted as
\begin{equation}
\vert N; 0^\eta ; (a, b);l^\pi;S; J^\pi M_J\rangle
\label{res_qnbr}
\end{equation}
Here, $\eta =\pm $ denotes the parity of the vacuum of the Fock space
accommodating the RS cluster, $(a, b)$= $\left( {K\over 2}, {K\over 2}\right)$,
$l$ is the underlying three-dimensional angular momentum of parity
$\eta (-1)^l$, $S$ is the quark spin, while $J^\pi $ and $M_J$ are in turn 
total spin and magnetic quantum numbers of the resonance under consideration.
In fact, $K$ is nothing but the four-dimensional angular momentum.

Within the framework of the rovibron model one can describe three
different types of transitions:

(i) Transitions without change of the quantum numbers $N$ and $K$,
i.e. transitions between resonances from same cluster.
In such a case, the transition operator is the $D_m'\, $ generator
of the $so(4)$ algebra and one can calculate the reduced probabilites
$B\left(\alpha_1,J_1\to \alpha_1 J_2; E1\right)$ for
 electric dipole transitions.
Notice that the reduced transition probability of the multipolarity $\lambda $
as carried out by the operator $T^{\alpha\, , \lambda }$
between states of initial and final spins $J_1$ and $J_2$, respectively,
is defined as \cite{Hey}
\begin{equation}
B\left( 
\alpha_1, J_1 \to 
\alpha_2, J_2; T^{ \alpha ,\lambda } 
\right)=
{1\over {2J_1 +1}}
\Big\vert
\left(
\alpha_2 J_2 
\vert\vert T^{\alpha, \lambda }
\vert\vert 
\alpha_1 J_1
\right)
\Big\vert ^2\, .
\label{B_lambda}
\end{equation}

Unfortunately, such transitions are difficult and perhaps even  
beyond any possibility of being observed.

(ii) Transitions between states of same number of bosons $N$ but 
of  different four dimensional angular momenta, $\Delta K\not= 0$, 
i.e. transitions between resonances
belonging to different RS clusters. 
Operators that can realize such transitions 
between different $O(4)$ multiplets 
are $U(4)$ generators (or, tensor products of them) 
lying outside of the  $so(4)$ sub-algebra.
The latter operators constitute the set 
\begin{eqnarray}
Q_m = \lbrack p^+\otimes \tilde{p}\rbrack^2_m\, , &\quad& 
E_m = {1\over \sqrt{2}}D_m\, ,
\nonumber\\
E0  &=& {1\over {2\sqrt{3}} }(3n_s -n_p)\, .
\label{elm_set}
\end{eqnarray}
It is not difficult to prove that the nine operators in
Eq.~(\ref{elm_set}) behave with respect to $SO(4)$ transformation
as the components of the totally symmetric rank-2 tensor,
$T^{(1,1)lm}$ where
\begin{eqnarray}
T^{(1,1)\, 2m}:=Q_m\, ,&\quad&
T^{(1,1)\, 1m}:=E_m\, ,\nonumber\\
T^{(1,1)\, 00}:&=& E0\, .
\label{11_tensor}
\end{eqnarray} 
By the way, the tensor $T^{(1,1)lm }$ is the one of lowest rank that can 
realize transitions between $SO(4)$ multiplets having same number of 
bosons $N$ and differing by two units in $K$.

iii) Transitions between $U(4)$ multiplets whose number of bosons differ
by one unit ($\Delta N=1$), the most interesting being resonance 
de-excitation modes into the nucleon
\begin{equation}
\vert N_1=5;0^{\eta} ;K_1;L_1;S_1={1\over 2}; J_1M_1\rangle \to
\vert N_2=4;0^{+} ;K_2=0;L_2=0;S_2={1\over 2}; {1\over 2}m_{{1\over 2}}\rangle 
\label{N_N_1}
\end{equation}
In the following we will be mainly interested in transitions of the 
third type.
At the present stage, however, it is convenient to first outline
the general scheme of the $SO(4)$ Racah algebra. 

Tensor products  
$\left[
T^{( a_1,b_2 )}  \otimes T^{(a_2,b_2)} \right]^{\left( a,b\right) 
lm } $ in $SO(4)$ are defined as
(see Refs.~\cite{Wy,Fi} for details)
\begin{eqnarray}
\left[T^{(a_2,b_2)} \otimes T^{ (a_1,b_1)}
\right]^{\left( a, b \right)  lm } 
=\sum_{l_1m_1 l_2m_2 
}&&\nonumber\\
\Big( \left( a_1b_1\right) \, l_1m_1\, \left( a_2b_2\right) \,
l_2m_2 \Big\vert
\left( a_1b_1\right)\, \left( a_2b_2\right)\, ;
\left( a b\right) lm \Big)
&& 
T^{\left( a_1,b_1 \right)\, l_1m_1 }
T^{\left( a_2,b_2 \right)\, l_2m_2 }\, .
\label{Tensor_pr_o4}
\end{eqnarray}
The matrix elements of any tensor operator 
$T^{(a,b)lm}$ between $O(4)$ states are expressed as
\begin{eqnarray}
\left<\left( a_1,b_1\right) ;l_1m_1\vert T^{(a,b)lm}\vert
\left(a_2,b_2\right);l_2m_2\right> &=&
\Big(\left( a_2b_2\right) \, l_2m_2\, \left(  ab\right) \,
lm\Big\vert \left( a_2b_2\right)\left( a b\right);
\left(a_1b_1 \right) l_1m_1
\Big)\,\nonumber\\
&&
\left(
\left(a_1,b_1 \right)
\vert\vert\vert
T^{(a,b)}\vert\vert\vert \left( a_2,b_2\right)
\right)
\label{doubl_red_me}
\end{eqnarray}
The $SO(4)$ Clebsch-Gordan coefficients entering the last equation are
determined by
\begin{eqnarray}
\Big(\left( a_2b_2\right) \, l_2m_2\, \left( ab\right) \,
lm \Big\vert
\left( a_1b_1\right)\, \left(  a_2b_2\right) \, ;
\left( a_1 b_1\right) l_1m_1
\Big)\, &=&\nonumber\\
\sqrt{(2l_1+1)(2l_2+1)(2l+1)(2a+1)(2b+1)}&&
(-1)^{(l-m)}\left(
\begin{array}{ccc}
l_1&l_2&l\\
-m_1&-m_2&m
\end{array}
\right) 
\left\{
\begin{array}{ccc}
a_1&a_2&a\\
b_1&b_2&b\\
l_1&l_2&l
\end{array}
\right\}\, .
\label{SO4_CLGo}
\end{eqnarray}

The last equation shows that the ratios of the reduced probabilities of
electromagnetic transitions between resonances with different
$K$  quantum numbers are determined as ratios
of the squared $SO(4)$ Clebsch-Gordan coeffecients, as the triple barred
transition matrix elements cancel out.
As an example of that type of transitions let us consider the electromagnetic
de-excitations of the natural parity resonances with spins $3/2^-$ and $1/2^-$ 
from the first cluster to the nucleon. 
 Obviously, the relevant tensor operator in $SO(4) $ 
space is $T^{\left( {1\over 2}, {1\over 2}\right)lm}$. 
The latter should connect $U(4)$ states with different numbers of bosons
i.e. $\Delta N=1$. Therefore, it can be taken in the form 
\begin{equation}
T^{\left({1\over 2},{1\over 2} \right)1m } =p_m^+\, ,
\quad T^{\left({1\over 2},{1\over 2} \right)00 }= s^+ \, .
\label{T_1/2_1/2}
\end{equation}
Transitions of the above type can then be calculated by means
of ordinary Racah algebra 
in considering $\alpha_i :=N\left(a_i,b_i \right)=N\left(K_i/2,K_i/2\right)$ 
(with $i=1,2$) as an intrinsic quantum number 
according to:
\begin{eqnarray}
\left< \alpha_1, l_1;{1\over 2};
\, J^\pi M_J\vert
T^{ \alpha ,  l m} \vert
\alpha_2, 0;{1\over 2};{1\over 2}^+m_{{1\over 2}}\right> & =&
(-1)^{(J- M_J) }\nonumber\\
\left(
\begin{array}{ccc}
   J&l &{1\over 2}\\
-M_J&m &m_{{1\over 2}}
\end{array}
\right)&&
\left( \alpha_1,l_1;{1\over 2};J^\pi \vert\vert
T^{\alpha, l}\vert\vert \alpha_2;0;{1\over 2}^+\right)\, .
\label{Yu_1}
\end{eqnarray}
In order to express double barred matrix element in terms of
triple barred matrix elements, the following relations should be 
taken into account:
\begin{eqnarray}
\left(\alpha_1, l_1;{1\over 2};J^\pi \vert\vert
T^{\alpha, l}\vert\vert \alpha_2; 0^+; {1\over 2};{1\over 2}^+\right)
& =& \delta_{l_1 l}
\sqrt{ 2(2 J + 1)}
\left( \alpha_1,l\vert\vert  T^{\alpha ,l}\vert\vert \alpha_2, 0\right)
\, ,\nonumber\\
\left(N (a_1, b_1);l_1\vert\vert  T^{(a, b)l}
\vert\vert N' (a_2, b_2);l_2
\right) & =&
\sqrt{(2 l_1 + 1)(2 l_2 + 1)(2 l +1 )(2 a_1 +1)(2b_1 +1)}\nonumber\\
&&\left\{
\begin{array}{ccc}
a_2&  b_2&  l_2 \\
a &  b &  l  \\   
a_1&  b_1&  l_1 
\end{array}
\right\}
\left( (a_1, b_1)\vert\vert\vert T^{(a,b)}\vert\vert\vert(a_2 ,b_2)\right)\,  ,
\nonumber\\
\left\{
\begin{array}{ccc}
0&  0&  0\\  
a&  b&  l\\   
a_1& b_1& l_1
\end{array}
\right\}
&=&\delta_{a_1a} \delta_{b_1b} \delta_{l_1l}                     
{1\over \sqrt{ (2 l +1)(2 a +1)(2 b +1)}}\,.
\label{Yu_2}
\end{eqnarray}
In combining Eqs.~(\ref{Yu_1}) and (\ref{Yu_2}) results into
\begin{equation}
\Big\vert\left(N \left(a_1,b_1\right);l_1;{1\over 2};J^\pi\vert\vert
 T^{\left(a,b\right) l }\vert\vert
N'\left( 0,0\right);0^+;{1\over 2}; {1\over 2}^+\right)\Big\vert^2=
(2J+1)\Big\vert \left(N\left( a, a \right)\vert\vert\vert  
T^{\left( a,a\right) }
\vert\vert\vert N' \left( 0, 0\right) \right)\Big\vert ^2 \, .
\label{final_me}
\end{equation}

\subsection{Electric De-excitations of Resonances to the Nucleon }

Eqs.~(\ref{Yu_1})-(\ref{final_me}) can be applied to calculate the ratio
of, say, the electric dipole de-excitations
D$_{13}$(1520)$\to p+\gamma$, and S$_{11}$(1535)$\to p+\gamma$.
In this case $l_1^\pi=l^\pi =1^-$, $a_1=a={1\over 2}$, 
$b_1=b={1\over 2}$, and $J^\pi $ takes the two values
$J^\pi ={3\over 2}^-$, and ${1\over 2}^-$, respectively.

Substitution of the relevant quantum numbers into 
Eqs.~(\ref{Yu_1})-(\ref{Yu_2}) followed by a calculation
of the ratio of the squared  values of the $J^\pi ={3\over 2}^-$, and
$J^\pi={1\over 2}^-$ matrix elements yields the theoretical ratio 
of the electric dipole widths of interest,
$\Gamma_\gamma ^{\mbox{D}_{13} }$,  and 
$\Gamma_\gamma ^{\mbox{S}_{11}}$ of the
respective D$_{13}(1520)$ and S$_{11}(1535)$ states as
\begin{equation}
{\cal R}
^{\mbox{th} } = 
\left(
{ { \Gamma_\gamma ^{\mbox{D}_{13} } } 
\over
            { \Gamma_\gamma ^{\mbox{S}_{11} } }
}
\right) ^{\mbox{th}}=1\, .
\label{widths}
\end{equation}
In order to compare it to data, one may approximate the dipole widths 
with the total $\gamma $ widths and obtain their experimental values  
from the full widths and the branching ratios listed in 
\cite{Part}. The full widths of the D$_{13}(1520)$ and S$_{11}(1535)$ 
resonances are reported as 120 MeV and 150 MeV, respectively.
The D$_{13}$(1520)$\to p+\gamma $ branching ratio is reported
as 0.46-0.56\%, while the S$_{11}$(1535) takes values within the broader
range from 0.15\% to 0.35\%. 
The theoretical prediction corresponds to a S$_{11}$(1535)$\to p+\gamma $   
ratio of 0.35\% and lies thereby at the upper bound of the data range.
This ratio is in fact $J$-independent. It shows that the purely algebraic
description is insufficient to reproduce the electromagnetic
properties of the resonances in great detail. 
In that regard, further development of the model is needed with the aim 
to account for the internal diquark structure.

\begin{quote}
Remarkably, the internal structure of the diquark does not show up in the 
spectra, and seems to be less relevant for the gross features of the 
excitation modes.
At the vertex level, however, it will gain more importance.
The merit of the rovibron model is that there it can be treated as a 
correction rather than as a leading mechanism from the very beginning.
\end{quote}

One can further compare gamma-widths of resonances carrying different
internal $O(3)$ quantum numbers $l$. This effect is easiest to study on the
example of  the natural parity resonances from the ``missing'' 
rovibron clusters.
To be specific, we will compare the reduced probabilities
for the following two transitions:
\begin{eqnarray}
\vert 4; 0^+; \, \left(2,2\right);
1^-; {1\over 2}; {3\over 2}^- m_{{3\over 2}}\rangle 
&\stackrel{T^{(2,2)1m} }{\longrightarrow}& 
\vert 4;\, 0^+;  (0,0); 0^+; {1\over 2};\,  
{1\over 2}^+m_{{1\over 2}}\rangle\, ,\nonumber\\ 
\vert 4; 0^+; \, \left(2,2\right);
3^-; {1\over 2}; {5\over 2}^- m_{{5\over 2}}\rangle 
&\stackrel{T^{(2,2)3m}}{\longrightarrow} & 
\vert 4;\, 0^+;  (0,0); 0^+; {1\over 2};\,  
{1\over 2}^+m_{{1\over 2}}\rangle 
\label{Missing_trans}
\end{eqnarray}
The relevant transition operator is
\begin{equation}
T^{ \left(2,2\right) lm }=
\lbrack T^{ \left( 1,1\right)} \otimes  T^{\left( 1,1 \right)}  
 \rbrack ^{(2,2)lm}\, .
\label{dip_oct}
\end{equation}
Here, $l$ can take the values $l=0,1,2,3$, and $4$. The first of the
transitions in Eq.~(\ref{Missing_trans}) is governed by the
electric dipole operator $T^{\left(2,2\right) 1m }$, while the second
is controlled by the electric octupole $T^{\left(2,2\right) 3m }$.
We are going to calculate the ratio ${\cal R}_2$ of the quantities
\begin{equation}
{\cal R}_2 = {{B\left(\alpha_1,{5\over 2}^-\to \alpha_2,
{1\over 2}^+; T^{(2,2)1}\right)}\over {B\left(\alpha_1, {3\over 2}^-\to 
\alpha_2, {1\over 2}^+;  T^{(2,2)3}\right)}}
\label{ratio_2}
\end{equation}
Here
\begin{eqnarray}
B\left(\alpha_1, {3\over 2}^-\to \alpha_2, {1\over 2}^+;  
T^{(2,2)1}\right)&=& {1\over 4}
\Big\vert \left(4;0^+;(1,1);1^-;{1\over 2};{3\over 2}^-\vert\vert
\lbrack T^{\left(2,2\right) 1 }\otimes 1\!\!1\rbrack^{(2,2)1} 
\vert\vert 4;0^+;(0,0);0;{1\over 2};{1\over 2}^+
\right)\Big\vert^2 \, ,\nonumber\\
B\left(\alpha_1, {5\over 2}^-\to \alpha_2, {1\over 2}^+; T^{(2,2)3}\right)&=& 
{1\over 6}
\Big\vert \left(4;0^+;(1,1);3^-;{1\over 2};{5\over 2}^-\vert\vert
\lbrack T^{\left(2,2\right) 3 }\otimes 1\!\!1\rbrack^{(2,2)3} 
\vert\vert 4;0^+;(0,0);0;{1\over 2};{1\over 2}^+
\right)\Big\vert^2 \, .
\label{E1_E3}
\end{eqnarray}
Usage of Eq.~(\ref{final_me}) yields equal reduced probabilities for both
the dipole and octupole de-excitations and thereby the unit value
for ${\cal R}_2$. Thus, within this early version of the
rovibron model, a given RS cluster will have
a common partial ($\gamma +p $)- decay width, 
that is insensitive to its $O(3)$ spin content. 

A more interesting situation occurs 
in the case of  LAMPF clusters, such like 
$\vert 5; 0^-; \, \left({3\over 2},{3\over 2}\right);
2^-; {1\over 2}; {3\over 2}^- m_{{3\over 2}}\rangle$.
There, one encounters a 
{\it suppression of electromagnetic transitions to the nucleon\/}. 
Indeed, in the rigorous case of an ideal O(4) symmetry,
due to the unnatural parities of the nucleon
resonances with masses above 1535 MeV (and the $\Delta $
excitations with masses above 1700 MeV), transitions of the type
\begin{equation}
\vert 5; 0^-; \, \left({3\over 2},{3\over 2}\right);
2^-; {1\over 2}; {3\over 2}^- m_{{3\over 2}}\rangle
 \to 
\vert 4;\, 0^+;  (0,0); 0^+; {1\over 2};\,  
{1\over 2}^+m_{{1\over 2}}\rangle 
\label{sel_rule}
\end{equation}
can not proceed neither via electric $E\lambda$- 
nor via magnetic
$M\lambda $ multipoles (to be presented elsewhere). 
In the less rigid scenario of a violated $O(4)$ symmetry,
mixing between states of same parity and total spins but different
$K$'s may occur. For example, the above unnatural
parity spin-${3\over 2}^-$ resonance  from the
$K=3$ multiplet  may mix up with the
spin-${3\over 2}^-$ of natural $(l=1^-)$ from the $K'=2$ multiplet
\begin{eqnarray}
\vert J^\pi ={3\over 2}^- m_{{3\over 2}}\rangle &=&
\sqrt{1-\alpha ^2}\vert 5;0^-; \left({3\over 2},{3\over 2}\right); 
2^-; {1\over 2}; 
{3\over 2}^- m_{{3\over 2}}\rangle
\nonumber\\
&+&\alpha  \vert 4;0^+; \left(1,1 \right) , 1^-; 
{1\over 2};{3\over 2}^- m_{{3\over 2}}\rangle
\label{mix_ing}
\end{eqnarray}
{}For similar reasons, also a mixing with $K'=4$ states can take place.
Within this mixing scheme, unnatural parity resonances can be 
excited electrically via their natural parity component.
As long as the relevant transition operator for such transitions is 
$T^{
\left( { {K'}\over 2},{{K'}\over 2}\right)lm
}$, its matrix element
between the nucleon and the resonance of interest will be proportional
to the mixing parameter $\alpha $. To be specific, 
\begin{eqnarray}
\langle J^\pi ={3\over 2}^- m_{{3\over 2}}\vert
T^{
\left(1, 1 \right)1m }
\vert 4;0^+; (0,0); 0^+, {1\over 2}; {1\over 2}^+m_{{1\over 2}}\rangle&& 
\nonumber\\
=
\alpha \langle 4;0^+; (1,1); 1^-;{1\over 2}; {3\over 2}^-  
m_{{3\over 2}} \vert T^{\left(1,1 \right)1m }
\vert 4; (0,0); 0^+; {1\over 2}; {1\over 2}^+m_{{1\over 2}}\rangle \, . &&
\label{Kprime_0}
\end{eqnarray}
It is obvious from the last equation, that electric
excitations of the nucleon into the unnatural parity resonances  
will be suppressed by the factor of $\alpha ^2$.
At the present early stage of development of the quark rovibron model,
the mixing parameter $\alpha $ can not be calculated but has to be
considered as free and determined from data. 
A theoretical prediction for $\alpha$ would require more fundamental approach 
to the internal diquark dynamics. In case the $O(4)$ symmetry
is slightly violated, one may assume $\alpha $ to be same for all
cluster inhabitants and perform some calculations as to
what extent such states can be linked via electromagnetic transitions
to the nucleon. 

\section{Summary and Outlook}
\noindent
The results of the present study can be summarized as follows:
\begin{enumerate}
\item The present investigation communicated
an idea of how Lorentz representations
of the RS type can emerge as {\it fundamental\/} as well as {\it composite\/}
free particles of definite mass and indefinite spin within the context 
of a relativistic space-time treatment of the three Dirac-fermion system.
Though structureless RS particles have not been detected so far,
the $N$ and $\Delta $ spectra strongly indicate existence of 
{\it composite\/} RS fields.

\item  Excited light unflavored baryons
preferably exist as multi-resonance clusters
that are described in terms of RS multiplets such as
the (predominantly) observed LAMPF clusters $2_{2I,+}$, $4_{2I,-}$ and 
$6_{2I,-}$, and the ``missing'' clusters $3_{2I,+}$ and
$5_{2I, +}$.

\item The above RS clusters accommodate all the resonances
observed so far in the $\pi N$ decay  channel 
(up to the $\Delta \left(1600\right)$ state). The LAMPF data
constitute, therefore, an almost accomplished excitation mode 
in its own rights, as only 5 resonances are ``missing'' for the 
completeness of this structure.

\item We modeled composite RS fields within the 
framework of the quark rovibron model and constructed a 
Hamiltonian that fits the masses of the LAMPF clusters. 

\item In using that Hamiltonian we predicted,
from a different but the $SU(6)_{SF}\otimes O(3)_L$
perspective, the masses of two ``missing'' clusters of natural parity 
resonances,  in support of the TJNAF ``missing'' resonance search program 
\cite{Burkert}.
``Missing''  resonances under debate in the literature,
such like  P$_{11}$(1880)\cite{Svarc} and P$_{13}$(1910)\cite{Tit}
could neatly fit into the $\left( 2,2\right)\otimes
\lbrack({1\over 2},0)\oplus (0, {1\over 2})\rbrack $ RS cluster
at 1935 MeV in Table I. 

\item
We constructed electric transition operators,
outlined the essentials of the O(4) Racah algebra,
and calculated ratios of reduced probabilities of various
resonance de-excitations to the nucleon.   
We found the internal structure of the diquark to be of minor importance
for the gross features of the excitation modes.
At the vertex level, however, a point-like diquark was shown
to be insufficient to account for differences in the branching
ratios of resonances  from same cluster.
It is that place where the present early version of the qRVM  
model of baryon structure needs further improvements. 
Treating the internal structure of the diquark as a correction 
rather than as a leading mechanism from the very beginning
is a major merit of the quark rovibron model.
\end{enumerate}

\section{Acknowledgement}
Work supported by CONASyT Mexico.

\end{document}